\documentclass[a4paper]{article}

\usepackage{INTERSPEECH2018}

\title{Voices Obscured in Complex Environmental Settings (VOICES) corpus }
\name{Colleen Richey$^2$* and Maria A.Barrios$^1$*, Zeb Armstrong$^2$, Chris Bartels$^2$, Horacio Franco$^2$, Martin Graciarena$^2$, Aaron Lawson$^2$, Mahesh Kumar Nandwana$^2$, Allen Stauffer$^2$, Julien van Hout$^2$, Paul Gamble$^1$, Jeff Hetherly$^1$, Cory Stephenson$^1$, and Karl Ni$^1$}
\address{
  $^1$Lab41, In-Q-Tel Laboratories, Menlo Park, CA 94025\\
  $^2$SRI International, Menlo Park, CA 94025 \\
  * Equal author contribution
}
\email{colleen@speech.sri.com, mbarrios@iqt.org}

\begin{document}

\maketitle
 
\begin{abstract}
   This paper introduces the Voices Obscured In Complex Environmental Settings (VOICES) corpus, a freely available dataset under Creative Commons BY 4.0. This dataset will promote speech and signal processing research of speech recorded by far-field microphones in noisy room conditions. Publicly available speech corpora are mostly composed of isolated speech at close-range microphony. A typical approach to better represent realistic scenarios, is to convolve clean speech with noise and simulated room response for model training. Despite these efforts, model performance degrades when tested against uncurated speech in natural conditions. For this corpus, audio was recorded in furnished rooms with background noise played in conjunction with foreground speech selected from the LibriSpeech corpus. Multiple sessions were recorded in each room to accommodate for all foreground speech-background noise combinations. Audio was recorded using twelve microphones placed throughout the room, resulting in 120 hours of audio per microphone. This work is a multi-organizational effort led by SRI International and Lab41 with the intent to push forward state-of-the-art distant microphone approaches in signal processing and speech recognition.

\end{abstract}

\noindent\textbf{Index Terms}: corpus, speech recognition, speaker recognition, data collection, LibriSpeech

\section{Introduction}

SRI International and Lab41, In-Q-Tel, are proud to release the Voices Obscured in Complex Environmental Settings (VOICES) corpus, a collaborative effort that brings speech data in acoustically challenging reverberant environments to the researcher. Clean speech was recorded in rooms of different sizes, each having distinct room acoustic profiles, with background noise played concurrently. These recordings provides audio data that better represent real-use scenarios. The intended purpose of this corpus is to promote acoustic research including, but not limited to:

\begin{description}
    \item[Speech Processing] Speaker Identification, speech recognition, speaker detection
    \item[Audio Classification] Event and background classification, speech/non-speech
    \item[Acoustic Signal Processing] source separation and localization, noise reduction, general enhancement, acoustic quality metrics
\end{description}
The corpus contains the source audio, the retransmitted audio, orthographic transcriptions, and speaker labels. The ultimate goal of this corpus is to advance acoustic research by providing access to complex acoustic data. The corpus will be released as open source, Creative Commons BY 4.0, free for commercial, academic, and government use.

Datasets for speech research are typically expensive, limited in scope, and behind paywalls. Synthetic data can be created by superimposing audio samples from datasets of isolated speech and noise and using software to generate reverberation\cite{farfield2010}. Unfortunately, these techniques do not accurately represent the acoustics of real-world environments and dynamic noise. Datasets collected in real environments often use few speakers\cite{farfield2007}.    


Successfully deploying speech and acoustic signal processing algorithms in the field hinges on access to realistic data. To this end, audio for the VOICES corpus was recorded under realistic, noisy conditions, that better represent real-use situations. In the remainder of this paper, a detailed description of the VOICES corpus is provided, including model baselines for automatic speech recognition and speaker identification. Section~\ref{sec:collection} describes the collection effort itself, Section~\ref{sec:statistics} provides some insight into the statistics of the dataset, and Section~\ref{sec:baselines} outlines model baselines that were run on the dataset. The corpus will be available on Amazon Web Services, where details on use cases and a download link will be provided.

\section{Dataset Collection}
\label{sec:collection}
The main focus when developing the VOICES corpus was to provide an open-source dataset centered on distant microphone collection under realistic conditions. Pre-recorded foreground speech and background noise were played in two furnished rooms with different acoustic profiles (reverberation, HVAC background, echo, etc.) and recorded by 12 distant microphones. Recording rooms were windowed and carpeted, with mostly bare walls and bare ceiling, furnished with tables and chairs. Four recording sessions were held in each room: one for each distractor noise type (television, radio, or babble) played concurrently with the foreground speech, and one session with foreground speech only. One hour of only distractor noise or ambient room background noise was recorded at the end of each session. This resulted in over 120 hours of recorded speech per microphone, for a total of 374,688 audio files.   

\subsection{Audio Sources}

The audio for foreground speech and distractor noise was selected from sources either in the public domain or under a creative commons attribution license that permits data derivatives and commercial use.

\subsubsection{Foreground Speech}

A total of 15 hours (3,903 audio files) were selected from LibriSpeech\cite{LibriSpeech2015}, a corpus of audiobooks in the public domain. All audio contains English read speech. Audio was taken from 300 speakers in the "clean" data subsets, with an even split between females and males. At least three minutes of speech were selected from each speaker, with at least one minute from three different book chapters - an amount sufficient for speaker identification tasks. LibriSpeech files use a sample rate of 16kHz, 16-bit precision, and Free Lossless Audio Codec (FLAC) encoding. Selected files were corrected for DC offset, normalized based on their peak amplitude, and converted to WAV format. The selected audio files were concatenated together with 2 seconds of intervening silence into a continuous audio file. The loudspeaker playing the foreground speech was on a motorized rotating platform. The order of the individual audio files was randomized, to guarantee that there was no correlation between a particular human speaker and a position of the loudspeaker. Signals to evaluate the room response were added at the beginning of each session. These included a steady tone, a rising tone, and a transient noise. The final concatenated source file was 19 hours long. 

\begin{figure}[t]
  \centering
  \includegraphics[width=\linewidth]{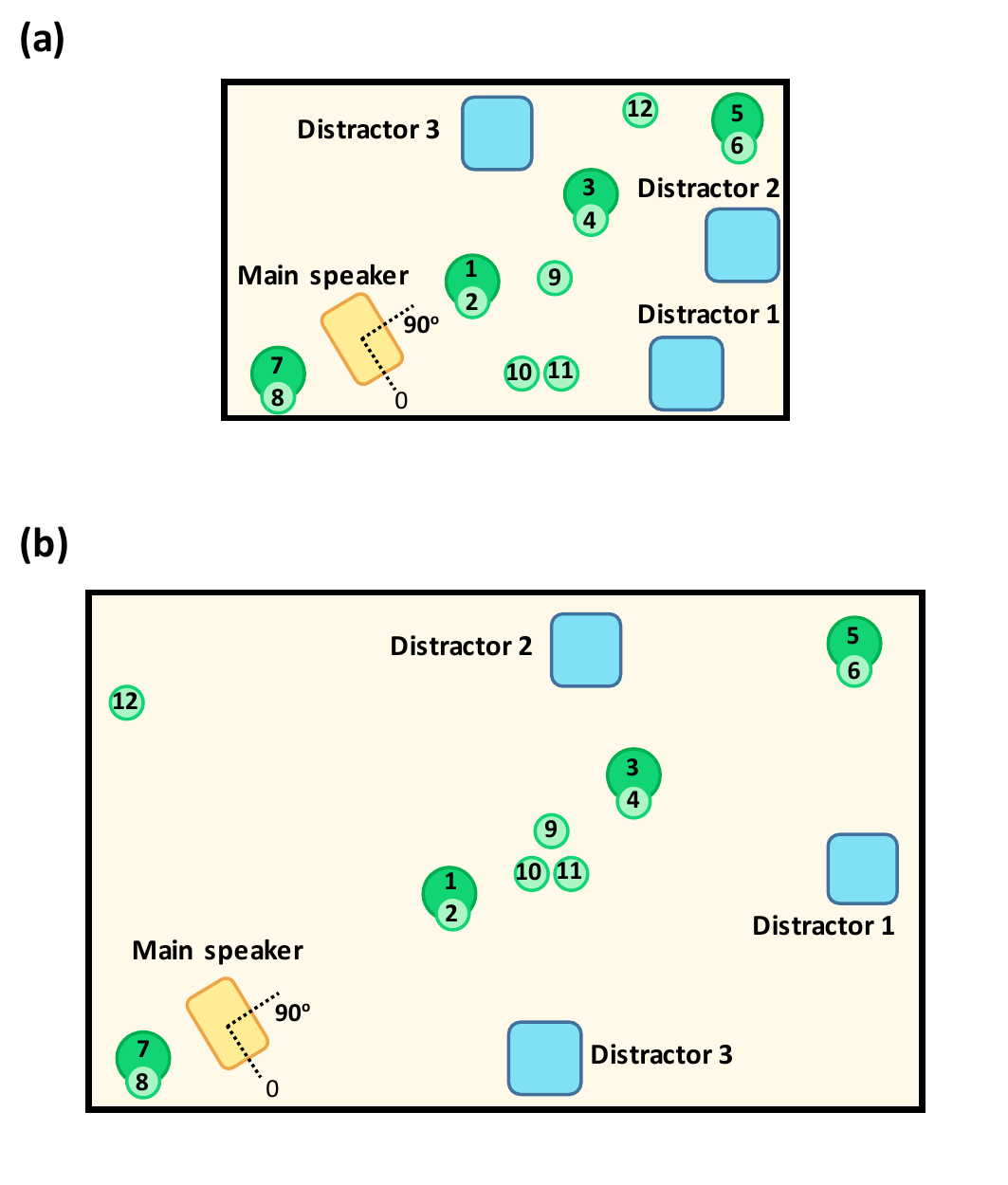}
  \caption{Microphone and loudspeaker configuration (not to scale) used for recording sessions in (a) room 1 (146" x 107") and (b) room 2 (225" x 158"). The foreground loudspeaker (shown here at its $90^\circ$ position), orange rectangle, was placed in a corner of the room, and speakers playing noise, blue squares, were placed with their cones directed toward the center of the room. Studio and lavalier microphones are shown as large (dark) and small (light) green circles; microphone number labeling corresponds to those outlined in Table~\ref{table:mics_dist}.}
  \label{fig:rooms_mics}
  \vspace{-5mm}
\end{figure}

\subsubsection{Distractor Noise}

Audio was recorded under four different noise conditions: one without any added noise (ambient room noise only) and three with a distractor noise played simultaneously with the foreground speech. The distractor noises were television, music, or overlapping speech from multiple speakers (referred to here as babble). During recording sessions, the audio for television or music was played from a single loudspeaker; babble was played from three noise-dedicated loudspeakers. An extra hour of just distractor noise was recorded at the end of each session. 

Television noise was selected from movies and television shows in the public domain\cite{PublicMoviesNet,PublicMoviesOrg}. Audio from 76 videos was extracted in M4A format and converted to WAV with a 16kHz sample rate and 16-bit precision. Five-minute excerpts were chosen from each audio file and each excerpt was normalized to its peak amplitude. Depending on the length of the source audio, 5 to 8 excerpts were taken from each movie or show, randomized, and concatenated into a single 20-hour audio file.

Music noise was selected from the MUSAN corpus\cite{MUSAN2015}.
All music files are in the public domain or under a Creative Commons license. Any music files having no derivative (ND) or non-commercial (NC) license restrictions were omitted from the sample set. The music files were randomized and concatenated into a single 20-hour audio file. Due to the large variability in signal amplitudes for different genres of music, the concatenated audio file was run through the compander tool in the SoX audio utility, combining compression and expansion of the signal dynamic range. This ensured a more uniform music volume throughout the recording sessions and a more consistent signal-to-noise ratio.

Babble noise was constructed using the "us-gov" subset of the MUSAN corpus\cite{MUSAN2015}. This subset contains audio recording excerpts of various US government meetings; all are in the public domain. Each excerpt is about 5 minutes long and was normalized to its peak amplitude. Babble tracks were constructed by randomizing and concatenating together meeting excerpts into 20-hour audio files and then mixing three audio files into one. Three babble tracks were created and were played out of three noise-dedicated loudspeakers (i.e., nine overlapping voices) simultaneously with the foreground speech. 

\subsection{Recording Setup}

Two different rooms were used for recording: room-1 with dimensions 146" x 107" (x 107" height) and room-2 with dimensions 225" x 158" (x 109" height). Twelve microphones were placed in strategic locations throughout the room: 7 cardioid dynamic studio microphones (SHURE SM58), 4 omnidirectional condenser lavalier microphones (AKG 417L), and 1 omnidirectional dynamic lavalier microphone (SHURE SM11). Paired studio and lavalier microphones were placed at four different positions: (1) Behind the foreground loudspeaker, (2) on a table directly in front of the foreground loudspeaker, (3) on a table in front of the foreground loudspeaker at a farther distance than (2), and (4) across the room from the foreground loudspeaker. The remaining four lavalier microphones were placed in other locations in the room, fully or partially obstructed by a physical barrier. Distances between the foreground loudspeaker and microphones are listed in Table~\ref{table:mics_dist}. All audio was played on high-quality speakers; one speaker was reserved for foreground speech, and three others were used to play distractor noise. A schematic of speaker and microphone placement in both rooms in shown in Figure~\ref{fig:rooms_mics}. 

\begin{table*}[t]
  \caption{Microphone type, location, distance from foreground loudspeaker (\textit{s}) and height (\textit{h}) for room-1 and -2 configurations.}
  \label{table:mics_dist}
  \centering
  \begin{tabular}{llcc} 
    \toprule
    \textbf{Mic ID (type)}  &\textbf{Location}  &\textbf{Room-1 (\textit{s}, \textit{h})} &\textbf{Room-2 (\textit{s}, \textit{h})}  \\
    \midrule
    01 (studio), 02 (lavalier) & near on table & (38", 42") & (80", 39")    \\
    03 (studio), 04 (lavalier) & far on table & (72", 42") & (131", 39")\\
    05 (studio), 06 (lavalier) & across room & (119", 70") & (228", 70") \\
    07 (studio), 08 (lavalier) & behind loudspeaker & (29", 70") & (29", 70")\\
    09 (lavalier)   & partially obstructed, table & (58", 28") & (109", 25")\\ 
    10 (lavalier)   & on ceiling, clear & (75", 105") & (128", 105")\\ 
    11 (lavalier)   & on ceiling, fully obstructed & (75", 106") & (128", 106") \\ 
    12 (lavalier)   & fully obstructed, wall & (130", 12") & (116", 10")\\ 
\bottomrule
  \end{tabular}
\end{table*}

The foreground speaker was placed 43" from the floor on a robotic platform that automatically rotated the position of the foreground speaker by ten degrees every hour, spanning a total of 180 degrees. The rotating platform's step motor was sufficiently shielded to prevent recording background noise from the motor movement. The motivation to have a non-static audio source was to emulate common human behavior that occurs during conversations such as head movement or walking, that is not captured in other datasets. 

A PreSonus StudioLive RML32AI digital mixer and PreSonus Capture recording software were used to play and record the audio. A sound pressure meter, placed close to microphone 01, was used to measure the playback audio and adjust volume levels on the PreSonus mixer for both the foreground audio (\textasciitilde65 dB) and distractor noise (\textasciitilde50 dB). All channels were sample synchronous. Each recording session lasted 20 hours (19 hours of foreground speech and 1 hour of only distractor or ambient noise). The recording sessions were segmented according to the source files from LibriSpeech, yielding 1440 hours of audio (347,688 audio files) across all microphones and sessions. Audio was recorded with a 48kHz sample rate and 24-bit precision in WAV format with PCM encoding, and is also available in 16kHz and 16-bit precision in WAV format. The corpus also contains the source audio files (16kHZ sample rate, 16-bit precision, WAV format).
 
\section{Data Statistics}
\label{sec:statistics}

To obtain an assessment of the statistics of the corpus, the duration, minimum and maximum amplitude, root mean square (RMS) energy, and signal-to-noise ratio (SNR) were calculated for all audio files in the corpus. Statistics were calculated using a combination of the SoX utility and SRI's in-house utilities. 

The average and median duration for all data subsets is 15.62s and 15.97s, respectively, with a standard deviation of 1.91s. This is evidence that the automatic audio segmentation worked correctly and that we can directly compare noisy files with source files. 

The RMS, measuring the amplitude of the audio file relative to the digital system's maximum level (with maximum value at 0 decibels relative to full scale - dBFS), was consistent across the various subsets. Average values were measured between -27.0 and -27.5 dBFS, indicating the playback volume was consistently set for all recordings. 

The minimum and maximum amplitudes represent the lowest and highest amplitude for samples in a given audio file, on a normalized scale of $\pm 1$. These were measured to be between -0.5 to 0.5 across all data subsets, showing reasonable use of the digital recording system’s levels. The average minimum and maximum amplitude levels for the source audio were -0.93 and 0.91. 

The SNR measures the strength of a primary signal relative to the background noise. Differences in SNR were evident between rooms and distractor noises, and in general degraded with increasing distance between the foreground loudspeaker and microphone. The average SNR for audio recorded in room-1 and room-2 was 22.19 dB and 19.50 dB, respectively. 

Table~\ref{table:SNR} shows the calculated SNR for audio recorded under different noise conditions as compared to the source audio's SNR. The SNR significantly degrades for audio recorded at a distance in a real acoustic environment, even without distractor noise. A decrease of 18 dB was observed for this case. The addition of noise further decreases the SNR, the worst is with babble noise. The SNR for microphones close to and behind the foreground loudspeaker was 22.3 dB, and for those at mid- and far-distance, it was 20.5 dB.

\begin{table}[h]
	\centering
	\vspace{2mm}
	\fontsize{8.5}{8.5}\selectfont
	\caption{Measured signal-to-noise ratio for the source audio and audio recorded at a distance with and without distractor noise.}
	\label{table:SNR}
	\begin{tabular}{c c c c c c}
		\toprule
		 & Source & No distractor & Music & TV & Babble\\
		\midrule
		SNR & 41.7 & 23.6 & 19.2 & 22.2 & 18.4\\
		\bottomrule
	\end{tabular}
	\vspace{2mm}
\end{table}

\section{Model Baselines}
\label{sec:baselines}
SRI's in-house automatic speech recognition (ASR) and speaker identification (SID) systems were used to examine the recorded data. This provides data validation for analytics and a point of reference for future model implementations. 

\subsection{Automatic speech recognition (ASR)}
The ASR system was run on a subset of data: audio from lavalier microphones when the foreground loudspeaker was positioned at $90^\circ$ (directly aligned with microphones on table). The ASR system was built using the Kaldi Speech Recognition Toolkit \cite{Povey_ASRU2011}. It uses filterbank features and a time delay neural network (TDNN) and was trained on 500 hours of segmented English speech, which included data collected under DARPA's Translation Systems for Tactical Use (TRANSTAC) program and SRI proprietary data. Training audio is included twice, once in its original form and a second with artificially added reverberation. Because no full test or development subset of data from LibriSpeech is included in the VOICES corpus, a direct comparison with published ASR results using LibriSpeech is not possible. It is possible, however, to make a rough comparison with results using the dev-clean LibriSpeech dataset. Published results for this subset achieved 4.9\% and 7.8\% word error rate (WER), for models trained on LibriSpeech and on the Wall Street Journal data, respectively\cite{LibriSpeech2015}. The SRI system achieved a 9.3\% WER. 

Table \ref{table:WER} shows the WER when the foreground speaker is at $90^\circ$ (centered) as a function of distractor noise. Results show a sharp increase in WER for data recorded in realistic acoustic environments. The WER for audio recorded by distance microphones with no added distractor noise is 19.0\% - more than double the WER on the source audio. Added distractor noise degrade the performance further. The worst performance is on audio with babble noise, as this type of noise contains only speech and easily confuses the ASR system.

\begin{table}[h]
	\centering
	\vspace{5mm}
	\fontsize{8.5}{8.5}\selectfont
	\caption{WER as a function of distractor noise type for room-1 with foreground loudspeaker centered at $90^\circ$, obtained from in-house SRI ASR system.}
	\label{table:WER}
	\begin{tabular}{c c c c c c}
		\toprule
		 & Source & No distractor & TV & Music & Babble\\
		\midrule
		WER & 9.3 & 19.0 & 27.6 & 29.3 & 33.0\\
		\bottomrule
	\end{tabular}
	\vspace{5mm}
\end{table}

\begin{figure}[h]
  \centering
  \includegraphics[width=\linewidth]{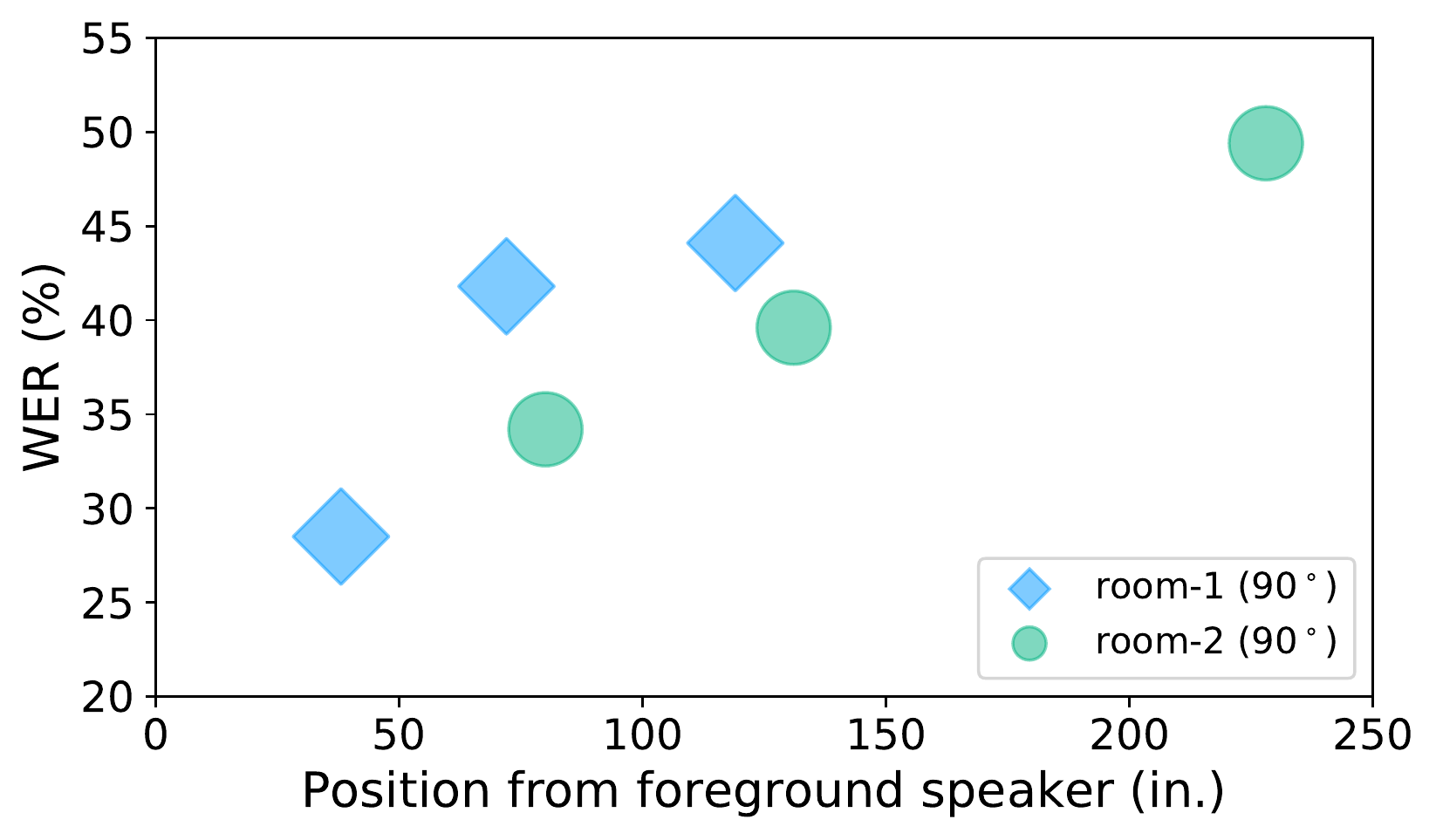}
  \caption{The WER performance is affected by distance from the foreground loudspeaker, as well as room acoustic profile.}
  \label{fig:WER_dist}
\end{figure}

In general the ASR performance is dependent on the distance between the foreground loudspeaker and microphone, and on individual room acoustics, as depicted in Figure \ref{fig:WER_dist}. Results are shown for microphones 02, 04, 06 in both rooms when the foreground loudspeaker is at $90^\circ$. There is an increase in WER with increased distance between the microphone and foreground loudspeaker. Differences in WER for microphones in room-1 and room-2 that are at comparable distances show the effect of each room's acoustic environment.   

\subsection{Speaker identification (SID)}
A state-of-the-art SID system from SRI was run on the VOICES corpus. The model used is a Universal Background Model (UBM) identity vector (i-vector) based system \cite{Dehak2011, GarciaRomero2011}, with a probabilistic linear discriminant analysis (PLDA) \cite{Prince2007} as back-end classifier. A gender-independent PLDA was used to compute the scores of the speaker recognition system. The model was trained using the PRISM dataset \cite{Ferrer2011}. The equal error rate (EER), describing the value where false positives equal false negatives, is used as the metric for the SID system performance. For our experimental setup, we ensured enroll and test audio segments corresponded to different book chapters from the original corpus. Speech segments were on average 14s long for both enroll and test subsets. Results are shown for microphones 01 and 02 (Close), microphones 03 and 04 (Mid), and microphones 05 and 06 (Far).

Table \ref{table:EER-distance} shows the impact of microphone distance on the SID performance. In this experiment, enrollment was performed on clean source data, and the EER is shown when testing on a variety of distant conditions. In order to highlight the effect of distance alone, no distractor noises were used. We observe that the EER of this UBM-IV system doubles when comparing the source audio ($5.72\%$) to audio from the close microphones for both rooms ($10.7\%$-$10.9\%$), and it almost triples for the far room microphone ($15.1\%$-$16.6\%$).

\begin{table}[h]
	\centering
	\vspace{5mm}
	\fontsize{8.5}{8.5}\selectfont
	\caption{Impact of microphone distance and placement on the performance of the UBM-IV speaker recognition systems in terms of EER (\%). Enrollment was done on source audio, and testing performed on distant microphones with no distractors.}
	\begin{tabular}{c c c c c}
		\toprule
		{\bf Mics} & Source & Close & Mid & Far  \\
		\midrule	
		\midrule
		Rm1 & 5.72 & 10.7 &13.0 &15.1 \\
		\midrule
		Rm2 & 5.72 & 10.9 &13.2 &16.6\\
		\bottomrule
	\end{tabular}
	\label{table:EER-distance}
	\vspace{2mm}
\end{table}

Table \ref{table:EER-distractor} shows the effect of distractor noise on SID performance. In order to mimic a realistic test case, speakers were enrolled using a recording from the close lavalier microphone (Close) in room-1 with no distractor noise. The test segments originate from all microphones and were recorded in room-2 with different types of background noise. We observe that distractor noise degrades the EER by $2\%$ absolute for music and television and $3.5\%$ absolute for babble. This is perhaps because it is very speech-like, but also possibly because babble was the only distractor played out of three separate loudspeakers.

\begin{table}[h]
	\centering
	\vspace{5mm}
	\fontsize{8.5}{8.5}\selectfont
	\caption{Impact of distractor noise on the performance of the UBM-IV speaker recognition systems in terms of EER (\%). Each condition has above $18k/2.8M$ target$/$impostor trials.}
	\begin{tabular}{c c c c c}
		\toprule
		{\bf Distractor} & No distractor & TV & Music & Babble\\
		\midrule
		UBM-IV & 17.2 &19.3 &19.2 &20.9\\
		\bottomrule
	\end{tabular}
	\vspace{5mm}
	\label{table:EER-distractor}
\end{table}

\section{Conclusions and Future Work}
The VOICES corpus provides audio data that closely resemble acoustic conditions found in real recording environments - distant microphones, background noise, and reverberant room acoustics. The corpus can serve as a test and development set for research in the areas of speech and acoustics. It will enable the development of robust acoustic models that can better perform in the wild. By making the corpus publicly available, SRI International and Lab41 hope to promote and advance acoustic research on event and background detection, source separation, speech enhancement, source distance and sound localization, speech activity detection, as well as speaker and speech recognition. Data presented here correspond to phase I data collection. The corpus will be augmented with further data collection in phase II, that will include additional rooms and more challenging distractor noise profiles.   

\bibliographystyle{IEEEtran}
\bibliography{mybib}

\end{document}